\documentclass[lettersize,onecolumn,journal]{IEEEtran}
\usepackage{hyperref}  
\usepackage{multirow}
\usepackage{booktabs}
\usepackage{comment}
\usepackage{hyperref}
\hypersetup{
    filecolor=blue,      
    urlcolor=blue,
    citecolor=cyan,
}
\usepackage{url}

\usepackage{cite}
\usepackage{colortbl}
\usepackage{color}
\definecolor{mygreen}{RGB}{232,231,239}
\newcommand*{\circled}[1]{\lower.7ex\hbox{\tikz\draw (0pt, 0pt)%
    circle (.5em) node {\makebox[1em][c]{\small #1}};}}

\usepackage{soul}

\ifCLASSINFOpdf
   \usepackage[pdftex]{graphicx}
\else

\fi

\usepackage{amsmath}
\usepackage{algorithmic}
\usepackage{array}

\newcommand{\etc}{\textit{etc}.}

\newcommand{\eg}{\textit{e}.\textit{g}.}

\begin{document}
\title{Perceptual Quality Assessment of Face Video Compression: A Benchmark and An Effective Method}

\author{Yixuan Li, 
        Bolin Chen, 
        Baoliang Chen, 
        Meng Wang, and  
        Shiqi Wang,~\IEEEmembership{Senior Member, IEEE},
        Weisi Lin, ~\IEEEmembership{Fellow, IEEE
        }
\IEEEcompsocitemizethanks{\IEEEcompsocthanksitem Y. Li, B. Chen, B. Chen, M. Wang, and S. Wang are with the Department of Computer Science, City University of Hong Kong, Kowloon, Hong Kong (e-mail: yixuanli423@gmail.com, bolinchen3-c@my.cityu.edu.hk, blchen6-c@my.cityu.edu.hk, mwang98-c@my.cityu.edu.hk, shiqwang@cityu.edu.hk).

W. Lin is with the School of Computer Engineering, Nanyang Technological University, Singapore (e-mail: wslin@ntu.edu.sg)

Corresponding author: Shiqi Wang.}\protect\\
}


\IEEEtitleabstractindextext{%
\begin{abstract}
Recent years have witnessed an exponential increase in the demand for face video compression, and the success of artificial intelligence has expanded the boundaries beyond traditional hybrid video coding. 
Generative coding approaches have been identified as promising alternatives with reasonable perceptual rate-distortion trade-offs, leveraging the statistical priors of face videos. However, the great diversity of distortion types in spatial and temporal domains, ranging from the traditional hybrid coding frameworks to generative models, present grand challenges in compressed face video quality assessment (VQA) that plays a crucial role in the 
whole delivery chain for quality monitoring and optimization.  
In this paper, we introduce the large-scale Compressed Face Video Quality Assessment (CFVQA) database, which is the first attempt to systematically understand the perceptual quality and diversified compression distortions in face videos. The database contains 3,240 compressed face video clips in multiple compression levels, which are derived from 135 source videos with diversified content using six representative video codecs, including two traditional methods based on hybrid coding frameworks, two end-to-end methods, and two generative methods. The unique characteristics of CFVQA, including large-scale, fine-grained, great content diversity, and cross-compression distortion types, make the benchmarking for existing image quality assessment (IQA) and VQA feasible and practical. 
The results reveal the weakness of existing IQA and VQA models, which challenge real-world face video applications. In addition, a  FAce VideO IntegeRity (FAVOR) index for face video compression was developed to measure the perceptual quality, considering the distinct content characteristics and temporal priors of the face videos. 
Experimental results exhibit its superior performance on the proposed CFVQA dataset. The benchmark is now made publicly available at: \url{https://github.com/Yixuan423/Compressed-Face-Videos-Quality-Assessment.}
\end{abstract}

}

\maketitle
\IEEEdisplaynontitleabstractindextext
\IEEEpeerreviewmaketitle

\section{Introduction}\label{sec:introduction}
\IEEEPARstart{F}{ace} video based services have been growing exponentially, coinciding with the accelerated proliferation of mobile communication and online video content sharing platforms. Face video compression towards human vision, which is indispensable in compressing and delivering gigantic-scale face video data, introduces visual distortions inevitably. During the past decade, advancements in video compression technology have quintessentially benefited face video compression. Face video coding schemes towards human vision include three types: traditional hybrid coding frameworks~\cite{zhang1992motion,h264,hevc,vvc}, convolutional neural networks (CNNs)-based frameworks~\cite{dvc,rlvc,deepc1,deepc2,deepc3,deepc4}, and generative adversarial networks (GANs)-based frameworks~\cite{lopez1995head,x2face,facev2v,fomm,oquab2021low,cfte}. To the best of our knowledge, while numerous codecs have been proposed, it appears that none have consistently achieved top-tier performance across all application scenarios, leaving great concerns for \textit{(1) how the quality varies among different face compression methods;} and \textit{(2) how to faithfully evaluate the quality of the compressed face videos with limited bit allocation.} 
\begin{figure}[tbp]
\centering
\includegraphics[scale=0.56]{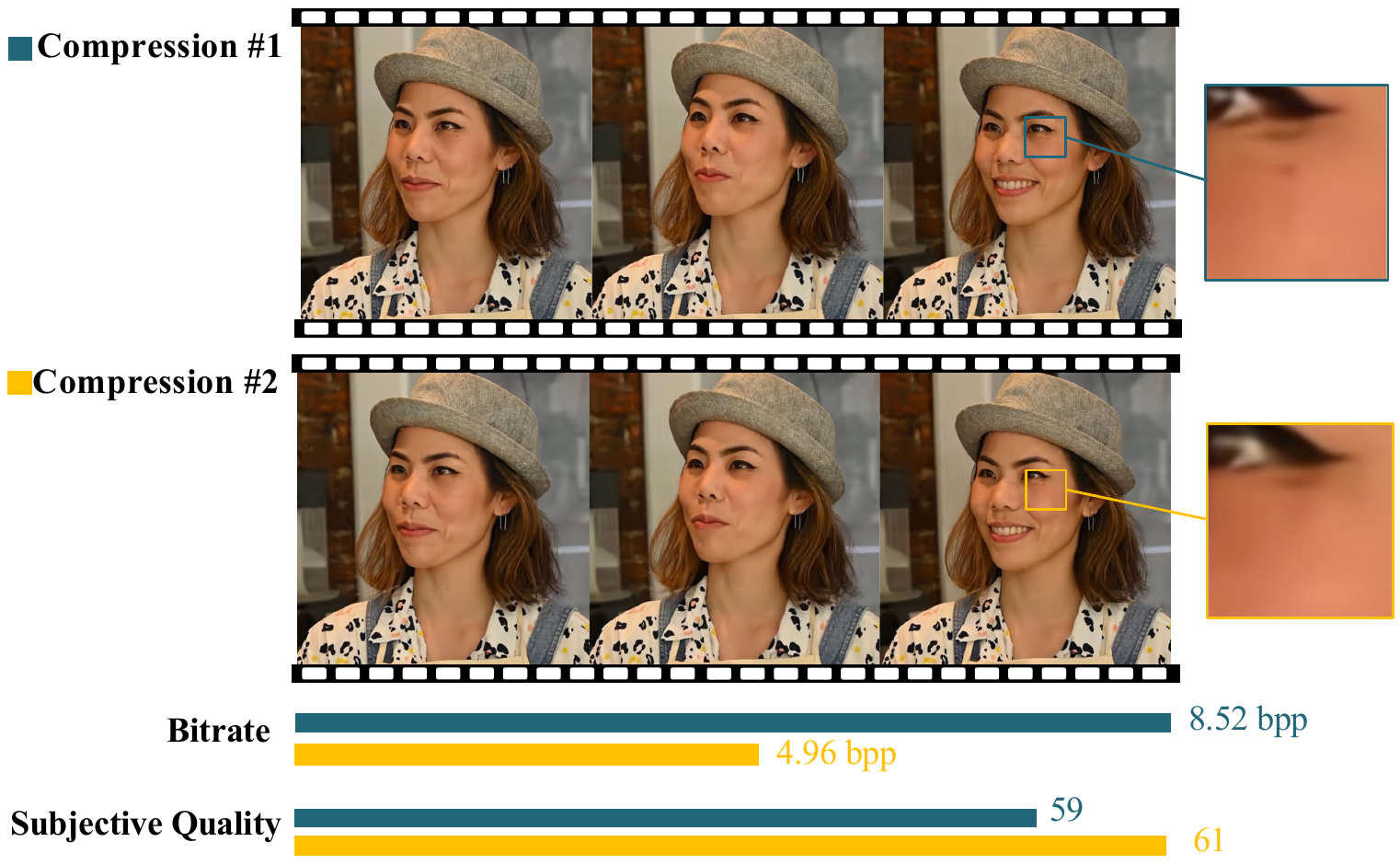}
\caption{Counterintuitive property of HVS quality perception on compressed face videos. Subjective quality is represented by the mean opinion score (MOS) rated by human subjects, and a higher value means a better quality.
Human vision indicates a significant preference for a ``good-looking'' face with appropriate smoothness caused by a higher compression level (lower bitrate). In comparison, the lower bitrate usually results in a lower quality for universal but non-facial content (\eg, buildings and scenery).  Video samples with such counterintuitive property are found common in our study, indicating the domain specialty of the quality assessment on compressed face videos.
}
\label{intro}
\end{figure}

For the first concern, pioneers have constructed various video quality assessment (VQA) datasets \cite{vqadb1,vqadb2,vqadb3}, which collect and evaluate compressed videos in various scenarios. However, these VQA datasets are typically designed for videos with generic content (\eg, indoor and outdoor scenes), VQA towards face video compression still requisites dedicated research. Unlike universal content, the human visual system (HVS) exhibits distinct perceptual characteristics when it comes to face video compression. As illustrated in Fig.~\ref{intro}, the video at a higher compression level is often preferred over one at a lower compression level, indicating the HVS tends to favor an appropriate level of facial smoothness over contrast enhancement. Moreover, geometric factors, such as facial symmetry and realism, significantly impact the quality of face videos. Motivated by the inherent differences, we here undertake a pathbreaking effort to create the Compressed Face Video Quality Assessment (CFVQA) dataset for facial VQA benchmarking and metric building. Compared with previous endeavors, CFVQA is the first large-scale VQA dataset specifically designed for compressed human faces. It encompasses a wide range of content and compression artifacts to cover the complete quality spectrum of face video coding used in practice. In particular, a total of 3,240 compressed face videos with more than 10,000 human ratings are included in our database.
The database boasts a high diversity of face video content (face, gender, age, ethnicity, face expression, scene, head position/motion, \etc) and distortion types (blur, blockiness, geometric distortions, flickering, floating, non-rigid motion, \etc). In addition to conventional hybrid coding schemes, we have included CNN-based and GAN-based coding schemes in our dataset, addressing the research gap in the quality assessment for cutting-edge face video compression models.

As for the second concern raised earlier, full-reference (FR) quality assessment measures, such as peak signal-to-noise ratio (PSNR), structural similarity index (SSIM)~\cite{ssim}, and video multimethod assessment fusion model (VMAF)~\cite{vmaf}, are extensively adopted to evaluate the objective quality of compressed videos. However, these measures do not consistently align with the human visual experience~\cite{liu2011image,xue2013gradient,ding2020image}, and often exhibit suboptimal performance when it comes to quality-distortion optimization in video compression. Recently, deep-learning-based IQA methods such as learned perceptual image patch similarity (LPIPS) \cite{lpips}, deep image structure and texture similarity (DISTS) \cite{ding2020image}, and DeepWSD ~\cite{liao2022deepwsd} have been proposed for a more accurate quality evaluation. Nevertheless, these methods were not specifically designed for evaluating compressed face videos and lack facial prior knowledge. In this paper, we propose a novel FR-VQA model for compressed face videos, where the prior knowledge of the human face is encoded in our quality-aware feature extractors and the frame-level quality is aggregated in the temporal domain with a proposed human memory-inspired module. The high performance of our proposed VQA model leads to a strong baseline for the future research on the quality assessment of compressed face videos. 

\section{RELATED WORK}

\begin{table*}[tbp]
  \centering
  \fontsize{7}{7}\selectfont
  \caption{Comparison with the previous VQA datasets in terms of reference video number, compressed video number, video content, and distortion type.}
    \setlength{\tabcolsep}{2mm}{\begin{tabular}{b{12em}<{\centering}p{6em}<{\centering}p{8em}<{\centering}p{5em}<{\centering}p{5em}<{\centering}p{28em}<{\centering}}
    \toprule
    \textbf{Dataset} & \textbf{\#Ref. videos} & \textbf{Video Content} & \textbf{\#Dist. videos} & \textbf{\#Dist. types} & \textbf{Dist. types}  \\
    \midrule
    EPFL-PoliMI\cite{de2009subjective,vqadb1}  &12  &universal &144 &1 &transmission packet loss \\
    \midrule
    LIVE-Mobile\cite{livemobile}    &10  &universal &200  &5  &\multicolumn{1}{m{28em}}{AVC SVC compression, transmission packet loss, frame-freeze, rate adaption, temporal dynamics} \\
    \midrule
    LIVE\cite{vqadb2}  &10  &universal  &150    &4  &\multicolumn{1}{m{28em}}{MPEG-2 and AVC compression, transmission packet loss in two modes}  \\
    \midrule
    IVP\cite{ivp}   &10  &universal    &128    &4  &\multicolumn{1}{m{28em}}{Dirac wavelet compression, AVC compression, transmission packet loss, MPEG-2 compression} \\
    \midrule
    IVC-IC\cite{ivc}    &60  &universal &240    &1  &AVC SVC compression \\
    \midrule
    TUM-HD\cite{vqadb3} &8 &universal &32 &1 &MPEG compression\\
    \midrule
    MCL-JCV\cite{vqadb4,vqadb5} &220    &universal  &45,760  &1 &AVC compression\\
    \midrule
    MCML-4K\cite{cheon2017subjective}   &10 &universal  &240   &3   &AVC, HEVC, and VP9 compression\\
    \midrule
    LIVE-Netflix\cite{bampis2017study}  &14 &universal  &112    &3  &\multicolumn{1}{m{25em}}{rebuffering events, dynamically changing AVC compression rates, mixture of both}    \\
    \midrule
    BVI-HD\cite{zhang2018bvi} &32   &universal &384  &1   &HEVC compression \\
    \midrule
    \textbf{CFVQA} (OURS)   &\textbf{135}    &\textbf{human face}  &\textbf{3,240}   &\textbf{6}  &\multicolumn{1}{m{28em}}{\textbf{VVC compression, reduced resolution compression, two E2E compression, and two generative compression algorithms}}\\
    \bottomrule
    \end{tabular}%
    }
  \label{tab:vqadatasetcompare}%
\end{table*}%
\subsection{Face Video Compression}
Dating back to the 1970s, Netravali and Stuller\cite{zhang1992motion} pioneered to propose a block-based motion compensation transform framework, which became the cornerstone of the hybrid prediction schemes broadly employed nowadays. Since the 2000s, video coding standards such as H.264/Advanced Video Coding (AVC)~\cite{h264}, H.265/High Efficiency Video Coding (HEVC)~\cite{hevc}, and H.266/VVC~\cite{vvc} have been developed and widely adopted for both academic and commercial applications. Recently, deep learning has boosted numerous data-driven algorithms~\cite{dvc,rlvc,deepc1,deepc2,deepc3,deepc4}, in particular the end-to-end (E2E) schemes that jointly optimize the compression process as a whole. For example, Lu \emph{et al.}\cite{dvc} proposed the first E2E deep video compression framework (DVC) for competitive RD performance. Similarly, Yang \emph{et al.}\cite{rlvc} proposed a Recurrent Learned Video Compression (RLVC) framework to leverage temporal correlation among video frames and improve the RD performance.

Furthermore, models dedicated to human face videos develops rapidly the huge proportion occupied in daily media streaming, where the statistical regularities which serve as the solid facial priors have motivated a series of face video compression algorithms.  
The pioneering compression scheme dedicated to the human face was proposed in \cite{lopez1995head}, where the 3D head model was adopted. The deep generative models such as variational auto-encoding (VAE) and generative adversarial networks (GANs) have also been leveraged in face video compression. In particular, Wiles \emph{et al.}\cite{x2face} proposed the X2Face to generate human faces using audio and pose code, which can be further applied to face compression. 
The ultra-low bit-rate few-shot video-to-video synthesis~\cite{facev2v} scheme was proposed in 2019, which translates raw videos to the compact version via the key-point representations. The First Order Motion Model (FOMM)~\cite{fomm} was proposed for face image animation, based on which Oquab \emph{et al.} proposed \cite{oquab2021low} for mobile video chat scenarios. Moreover, Chen \emph{et al.}\cite{cfte} proposed an E2E talking-head video compression framework (CFTE) via compact feature learning, achieving high compression efficiency under ultra-low bitrate scenarios. These face-specific video coding methods targeting low bit-rate have been claimed to be efficient in providing much better visual quality at the same level of coding bits.  
\subsection{Quality Assessment of General-content Video Compression}
\subsubsection{Subjective Video Quality Assessment}
Subjective quality assessment is the most reliable and straightforward way to evaluate visual quality. In particular, the subjective video quality assessment datasets, which contain distorted videos with diverse content, distortion types and levels, could serve as the golden ground truth in fairly comparing and testing the performance of different objective quality assessment models. Multiple well-known datasets have been constructed for VQA, including authentic-distorted datasets and artificial-distorted datasets. In 2009, Simone \emph{et al.}~\cite{de2009subjective} proposed the first subjective VQA dataset targeting exploring the video quality degradation caused by transmission packet loss. Subsequently, several datasets \cite{vqadb1,livemobile,vqadb2,ivp,ivc,vqadb3,vqadb4,vqadb5,cheon2017subjective,bampis2017study,zhang2018bvi} focusing on compressed video quality were proposed, which significantly boost the evolution of FR objective measures. The summary of the aforementioned datasets are listed in Table~\ref{tab:vqadatasetcompare}. 

However, there are fundamental limitations for the particular task studied in this work. First, the content of the existing datasets is not restricted to an dedicated narrow scope solely with face videos that are deemed to have unique domain pattern. Second, the distortion types of existing datasets are conventional (e.g., H.264/AVC or H.265/HEVC compression), while the distortions generated by learning-based compression algorithms, including E2E codecs and generative compression schemes are not 
considered. Third, a large-scale dataset with diverse content is highly desired, while a typical issue of the existing datasets is the limited number of reference videos with human faces. These reflect that the quality assessment of compressed face videos lags far behind the development of face video compression codecs.

\subsubsection{Objective Video Quality Assessment}
The full-reference (FR) video quality assessment (VQA) and image quality assessment (IQA) algorithms which quantify media quality by measuring the similarity between the reference and distorted videos, have been widely applied in video compression performance evaluation and optimization~\cite{wang2019extended, cfte,dvc,rlvc}. 
The most popular method PSNR~\cite{psnr} has been widely criticized due to the low correlation with the HVS~\cite{blau2018perception}. The SSIM~\cite{ssim} and its variants\cite{msssim,sampat2009complex, wang2010information,zhang2011fsim, liu2011image,xue2013gradient} have been developed with the design philosophy that the HVS is excelling in extracting the structural information. 
Recently, deep-learning-based IQA methods like learned perceptual image patch similarity (LPIPS)\cite{lpips} and deep image structure and texture similarity (DISTS)\cite{ding2020image} quantify the similarity in the deep-feature domain. In~\cite{liao2022deepwsd}, the reference and distorted features are compared with the Wasserstein distance.

Video quality assessment involves the temporal domain which significantly influences perceptual quality. Therefore, a series of non-blind VQA approaches\cite{yang2007perceptual,movie,moorthy2010efficient,manasa2016optical,vmaf,bampis2018spatiotemporal,kim2018deep,kim2020dynamic} perform satisfactorily on the general-content video quality assessment task. For instance, in \cite{movie}, the authors developed the motion-based video integrity evaluation (MOVIE) based on the spatial frequency and orientation selectivity of the HVS. In \cite{moorthy2010efficient}, Moorthy \emph{et al.} proposed a motion-compensated structural similarity index (MC-SSIM) that balances the trade-off between motion information and computational cost.  In 2016, the video multimethod assessment fusion model (VMAF)~\cite{vmaf} was proposed and became a popular performance evaluation criterion for video codecs. Besides, reduced-reference VQA methods\cite{strred,ma2012reduced} also significantly contribute the quality evaluation of compressed videos. Although these VQA models achieve good correlations with the HVS, they are not specifically designed for face videos, leading to less  satisfactory performance. Such conflict leads us to rethink the effective and efficient perceptual quality evaluation methods for compressed face videos. 
\subsection{Quality Assessment of Face Videos}
The face image/ video quality assessment (FIQA/FVQA) is a long-standing research problem, typically referring to two research areas. The most-developed one is sourced from the biometrics community, targeting at investigating the recognizability of face images or videos as the input to face recognition systems. The pioneering work of this kind was proposed in \cite{fiqa1}, where face asymmetries due to lighting and face pose are quantified. Subsequently, numerous FIQA algorithms emerge by extracting quality-aware features in a handcrafted \cite{fiqa2,fiqa3,fiqa4,fiqa5} or  learning way \cite{fiqa6,fiqa7,fiqa8,fiqa9,fiqa10}. For instance, Zhang \emph{et al.} \cite{zhang2017illumination} propose a illumination quality dataset of face images and a deep-learning-based measure. Terhorst \emph{et al.} \cite{fiqa9} proposed the SER-FIQ which computes the Euclidean distance of the unsupervised quality-aware features as face quality measure. In 2021, Schlett \emph{et al.} \cite{fiqasurvey} provide a comprehensive survey on biometric FIQA, summarizing representative FIQA databases and methodologies. However, these methods are not dedicated to compressed face videos, although certain connections could exist. Due to the distinction of the ultimate receiver, biometric evaluation methods cannot be directly utilized for evaluating the perceptual quality of compressed face videos.
\begin{figure*}[!tbp]
\centering
\includegraphics[scale=1.26]{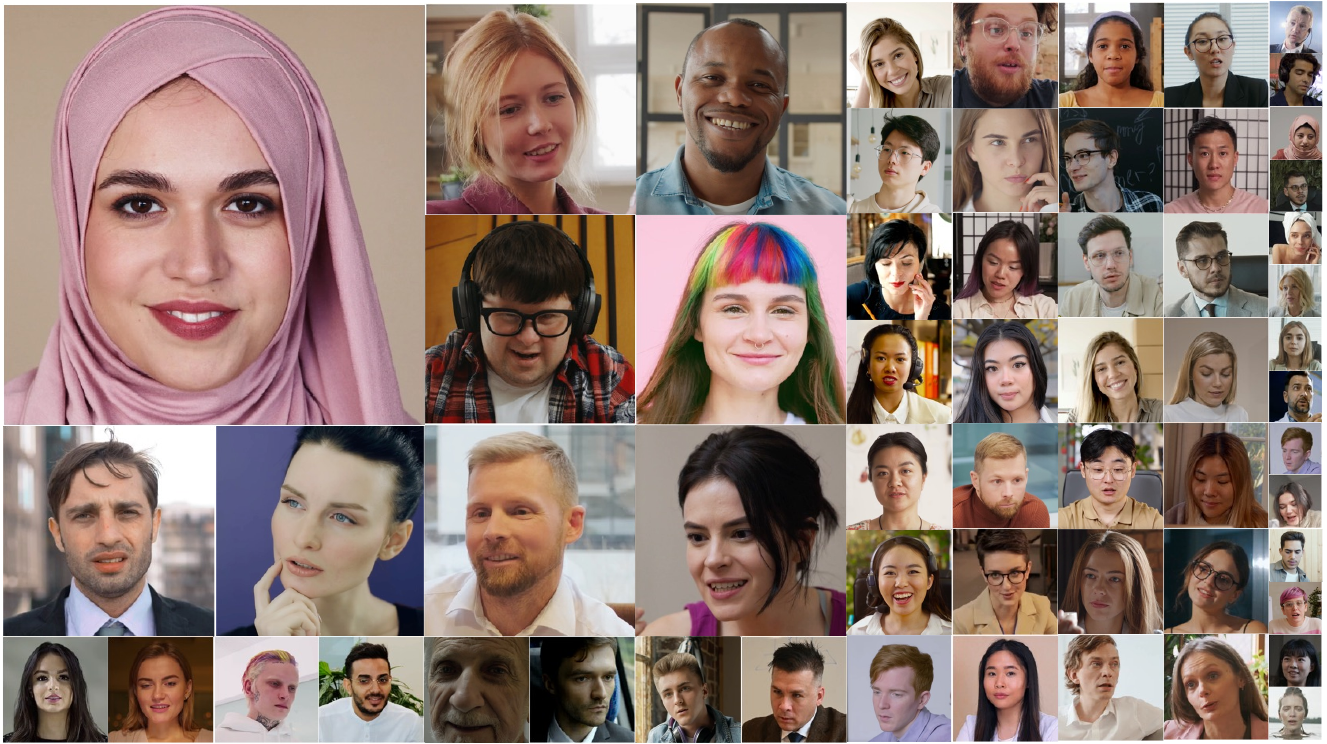}
\caption{Sample frames from reference videos in this benchmark. In total, there are 135 different reference face videos in the CFVQA, covering a great range of content diversity.}
\label{dataset-ref}
\end{figure*}

In contrast, another research line is assessing the perceptual quality of face images/videos in terms of Quality of Experience (QoE). 
In \cite{su2022going}, Su \emph{et al.} created the first IQA database of in-the-wild human faces called GFIQA-20k and proposed a blind FIQA model based on the generative face prior. In  \cite{jo2023ifqa}, Jo \emph{et al.} proposed another blind FIQA method assessing the perceptual realness of face images. 
However, both subjective and objective quality assessment for compressed face videos is still out of the research attention. Our elaborately designed dataset and the proposed model could mitigate such the gap and shed light on the high-quality face video compression.

\section{Constructing the CFVQA Dataset}\label{sec:database conrtuction}
\subsection{Face Video Content}
We construct the CFVQA dataset including 135 reference and 3,240 distorted face videos spanning a considerable diversity of face video content, in terms of various ages, ethnicity, appearances, expressions, head positions, head motion trails, camera motions, backgrounds, etc. The CFVQA also has good accessories coverage, like glasses, headsets, hijab, hairbands, etc.
The source face videos are obtained from two copyright-free websites, Pexel.com~\cite{pexels} and Mixkit.com~\cite{mixkit}, where the videos are permitted to use, modified, and redistributed freely without attribution. Considering the video coding scenarios, we crop the original videos into the resolution of $512 \times 512$, where the face area has the movement range limited in the cropped area. As suggested in \cite{videolength2,videolength3}, 5s-length sequences are truncated from each raw video, leading to 135 face videos with 16,875 frames in the resolution of $512 \times 512$, during for 5 seconds, and frame rate at 25 fps. In Fig.~\ref{dataset-ref}, we sample and illustrate part of the reference face video content. 

\subsection{Face Video Compression Models}\label{sec:compression}
We select six representative compression methods from mainstream face video compression codecs to generate compressed videos in the proposed benchmark, including two traditional methods, two CNN-based E2E methods, and two GAN-based generative methods. The descriptions of the codecs are as follows,
\begin{itemize}
    \item VVC~\cite{vvc}: Relying on the block-based hybrid coding framework.
    \item Reduced resoLution (RL):  Following the \emph{Downsampling, VVC compression and  Upsampling} workflow. The bicubic downsampling and upsampling are adopted. 
    \item DVC~\cite{dvc}: End-to-end framework with learning-based optical flow motion estimation module and auto-encoder style compression module.
    \item RLVC~\cite{rlvc}: End-to-end framework with recurrent auto-encoder and recurrent probability model.
    \item FOMM~\cite{fomm}: GAN-based model with keypoint representation.
    \item CFTE~\cite{cfte}: GAN-based model with deep learning based feature representation.
\end{itemize}

Each reference video is compressed by these codecs to three to five compression levels. Regarding the selection of the compression parameters (i.e., quantization parameters, QP), we follow two principles. First, the separation of different levels by adjusting the QPs enables a wide range of quality levels covered. Second, we ensure the alignment of the quality level for different compression methods by a small-scale subjective study. The detailed QP selection is detailed in the Supplementary Material (SM). In total, $135 \times 24=3,240$ compressed face videos are generated from 135 reference face videos. Herein, we provide thorough analyses of unique face video compression distortions in our benchmark.
\begin{figure*}[tbp]
\centering
\includegraphics[scale=0.58]{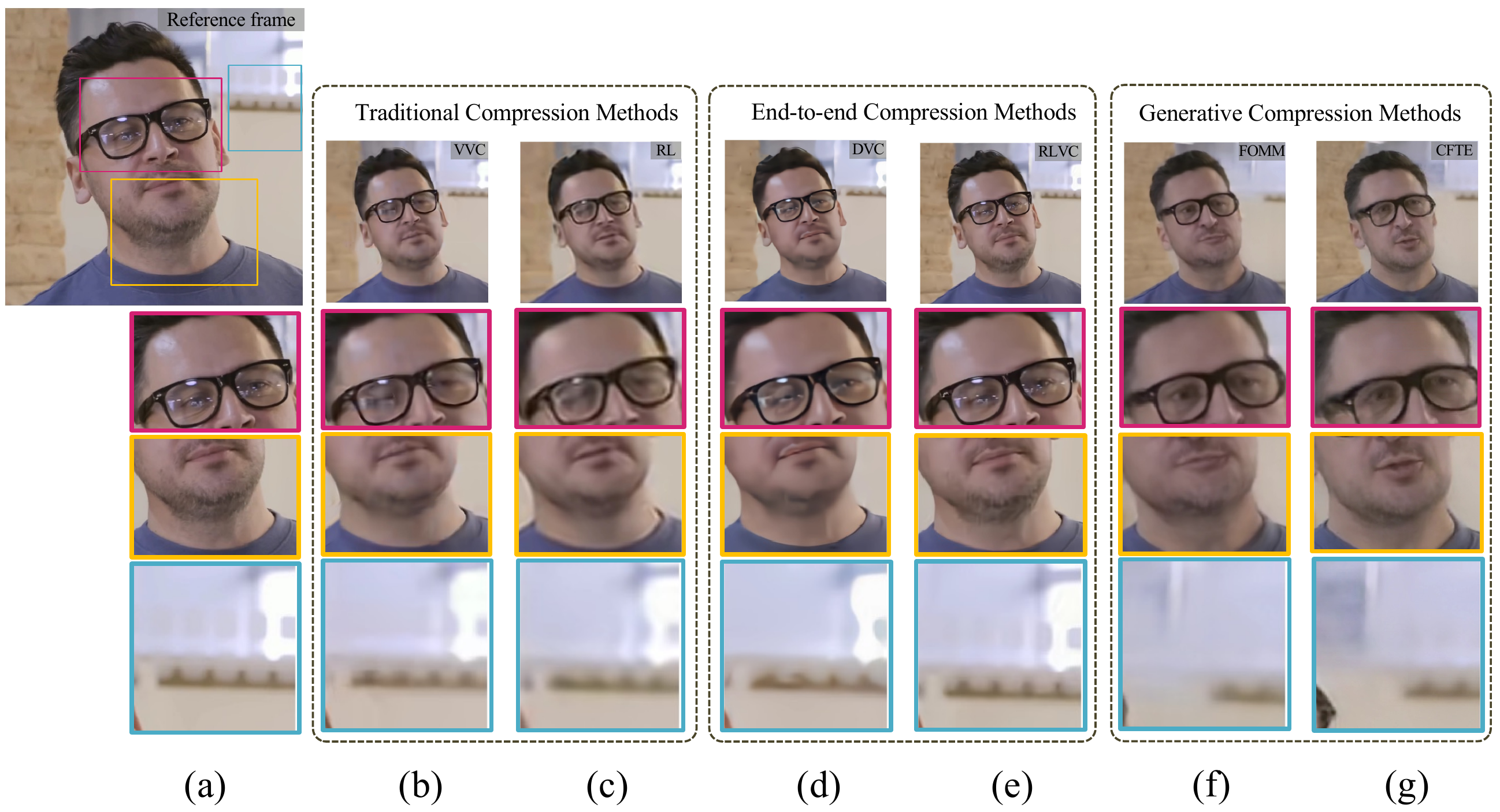}
\caption{Illustrations of spatial distortions in compressed face videos. (a) Reference frame and corresponding patches from the CFVQA; (b) VVC; (c) RL; (d) DVC; (e) RLVC; (f) FOMM; (g) CFTE. Zoom in for better vision.}
\label{compress-dist-s}
\end{figure*}

\begin{figure*}[t]
\centering
\includegraphics[scale=0.67]{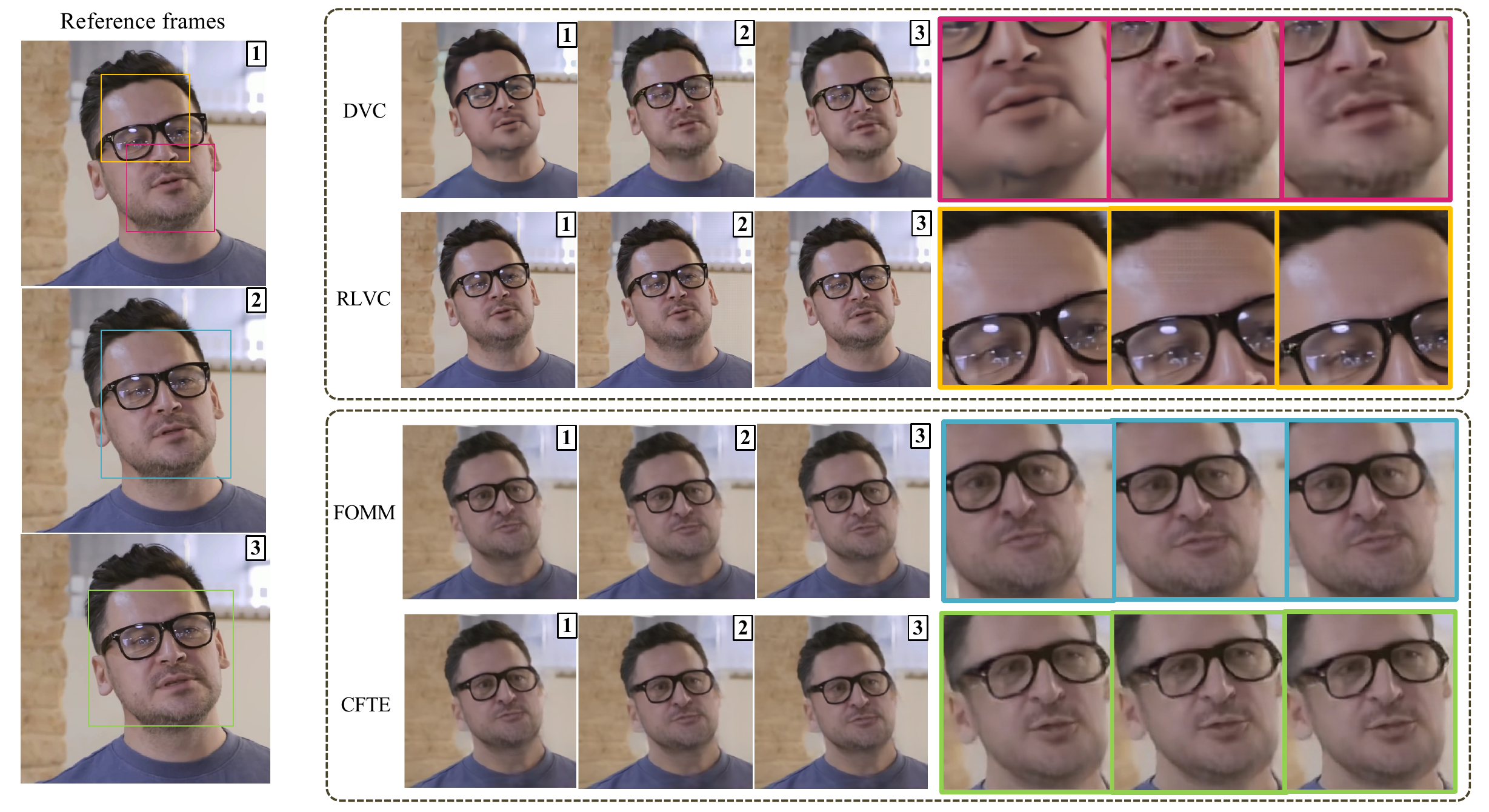}
\caption{Illustrations of perceptual temporal distortions in compressed face videos. Three successive frames are utilized for reference and distortion comparisons. The images are better viewed with zooming.}
\label{compress-dist-t}
\end{figure*}
\subsubsection{Spatial-domain Artifacts}
\begin{itemize}
    \item \textbf{Blur}. Blur has been the most commonly involved distortion type in modern video compression, which manifests in diversified forms in all six video compression methods, degrading video fidelity. However, in some occasions, blur does not necessarily degrade QoE of face videos but is regarded as pleasing smoothness in the E2E-compressed methods due to aesthetic factors.
    \item \textbf{Blockiness}. Blockiness is a typical compression distortion type in the traditional hybrid coding framework, causing visually disturbing discontinuities on the block boundaries. It is observed to appear in the face videos compressed by VVC and RL methods. 
    \item \textbf{Edge artifacts}. In this study, edge artifact is designated as widened edges and fake edges, which is caused by the resampling in the RL collection. For example, in Fig.~\ref{compress-dist-s}, fake edges can be found near normal edges in the RL-compressed video frames.
    \item \textbf{Geometric distortions}. Geometric distortion is comprised of content twisting and shape change in this study, which originates from generative models. Such artifacts often cause the sense of unreality of human faces, which is most eye-catching and annoying according to the user study in SM. Fig.~\ref{compress-dist-s} provides exemplar frames from CFTE and FOMM-compressed videos, where the face area is twisted weirdly, and some content is falsely eliminated.
    {\item \textbf{Deconvolution artifacts.} In Fig.~\ref{compress-dist-t}, checkboard-like patterns appear when one looks very closely at E2E compressed video frames (e.g. the forehead area), which is caused by the deconvolution operation when building up frames from low to high resolution in the encoder-decoder framework of E2E methods~\cite{odena2016deconvolution}. Such artifacts degrade both the perceptual quality and fidelity.}
\end{itemize}

\subsubsection{Temporal-domain Artifacts}
\begin{itemize}
    \item \textbf{Flickering}. Flickering artifact refers to frequent transients in terms of luminance, chrominance, and texture existing locally or globally in the video. 
    As shown in Fig.~\ref{compress-dist-t}, the texture of three consecutive frames compressed by DVC and RLVC codecs shows periodically variations, and such inconsistency is visually perceived as flashing effects.
    \item \textbf{Floating}. Floating artifacts refer to the illusive motions in certain regions as opposed to their surrounding background~\cite{compdist}. It often leads to abnormal jello-like content wobbling, manifested as either the stationary background areas surrounding face boundaries can even move along with the face, or certain areas may not vary along with the global camera motion. As a result of erroneous motion estimation in the generative coding methods, temporal floating always annoys visual experience. 
    \item \textbf{Non-rigid motions}. The non-rigid motion does not preserve the shape of objects, which is caused by the temporal generation inconsistency of the frames compressed by generative models. 
\end{itemize}

\subsection{Subjective Testing}
\begin{figure}[!tbp]
\centering
\includegraphics[scale=0.28]{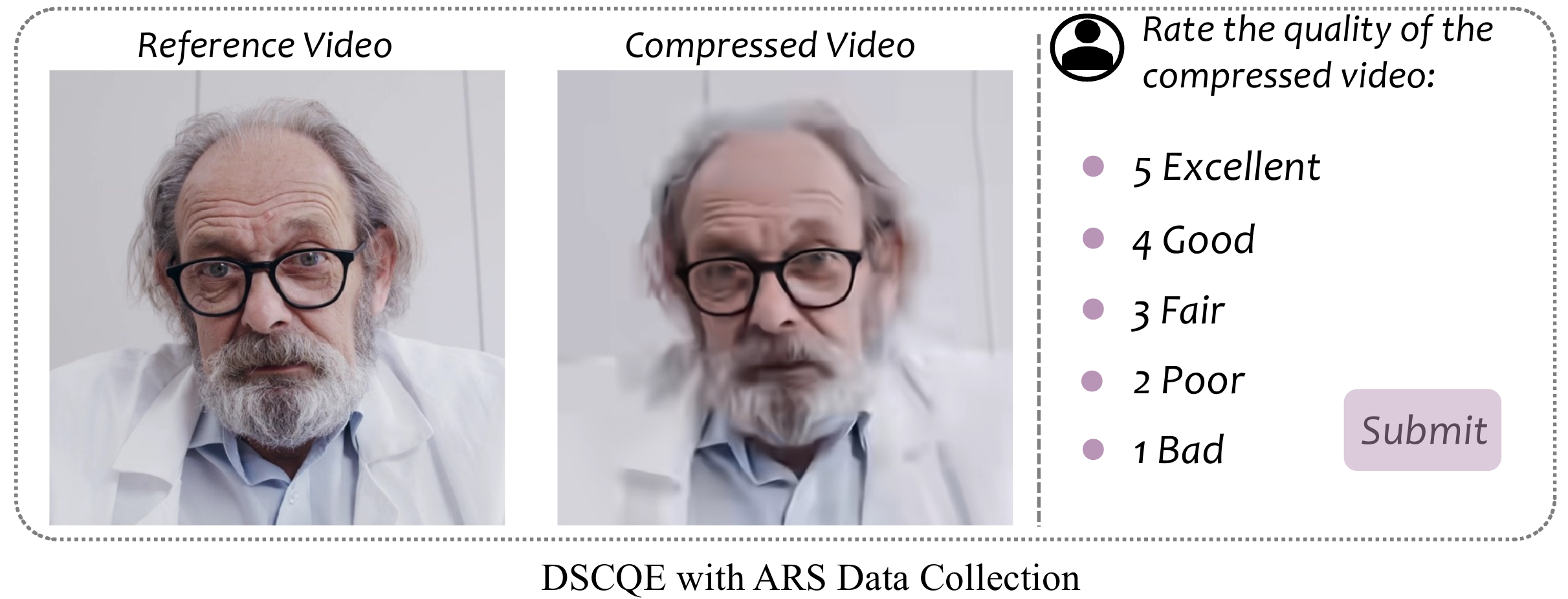}
\caption{Overview of subjective quality rating collection interface. The right side contains five labels for the users to rate quality adequately, including ``Excellent'', ``Good'', ``Fair'', ``Poor'', and ``Bad'' from the lowest level to the highest, denoting the quality scores ranging from 5 to 1.}
\label{subj-gui}
\end{figure}

Following ITU-R P.910\cite{ituvqa}, the subjective quality evaluation protocol is designed with the set-up testing environment. We adopt the Double-Stimulus Continuous Quality Evaluation (DSCQE)~\cite{dscqe} approach with the 5-category discrete Absolute Rating Scale (ARS)~\cite{ars} in our test. The user interface used in our test is shown in Fig.~\ref{subj-gui}.  Moreover, we take practical actions to avoid viewer fatigue, memory and contextual effect. In the end, a user study is conducted, aiming at investigating the perceptibility of different distortion types in compressed face videos. Following the recommendation regarding data reliability in \cite{ituvqa}, 32 subjects fully participated in the whole process of our study, from the age group of 22 to 30, and approximately half from each gender. The detailed settings and results of our subjective test can be found in the SM. 
\section{DATA PROCESSING AND ANALYSES}\label{sec:ANAlYSIS}
\subsection{Subjective Ratings Processing}\label{sec:mosgeneration}
To ensure the cognitive impenetrability (i.e., results in consistent decisions across different individuals for different video content), we conduct agreement testing and subject rejection as specified in the SM. Trustworthy subjective ratings are further converted into the MOSs following \cite{vqadb2}. Initially, the Z-scores are given by, 
\begin{equation}
Z_{ij} = \frac{S_{ij}-\mu_{i}}{\delta_{i}},
\end{equation}
where $i$ and $j$ denote subjective and test video indices. The $\mu_{i}$ and $\delta_{i}$ are the mean value and standard deviation on the quality rating $S_{ij}$ of subject $i$. Subsequently, Z-scores are linearly mapped to the range of $[0,100]$ under the assumption that Z-scores of one subject follow the standard Gaussian\cite{zscore} in the range of $[-3,+3]$,
\begin{equation}
{\tilde{Z}}_{ij} =\frac{100(Z_{ij}+3)}{6}.
\end{equation}
Finally, the MOS and corresponding standard deviation can be computed as
\begin{equation}
\omega_{j}=\frac{1}{N}\sum_{i=1}^{N}{\tilde{Z}}_{ij},
\end{equation}
\begin{equation}
\sigma_{j}=\sqrt{\frac{1}{N-1}\sum_{i=1}^{N}{(\tilde{Z}}_{ij}-\omega_{j})^2},
\end{equation}
where $N$ is the number of subjects involved.

\begin{figure}[!tbp]
\centering 
\includegraphics[scale=0.33]{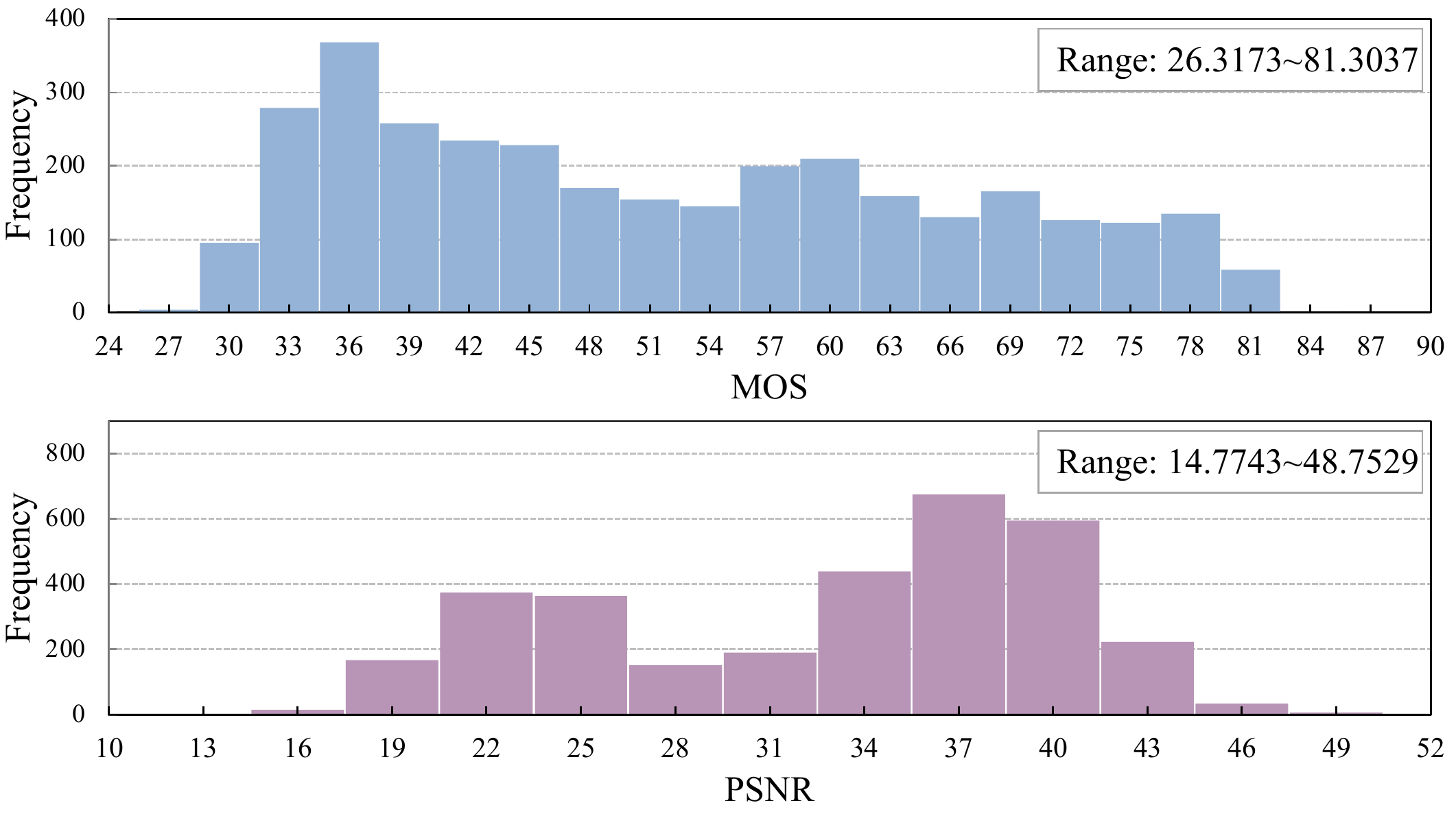}
\caption{Distribution of MOSs, and PSNR of all the videos in the proposed CFVQA benchmark.}
\label{mos}
\end{figure}

The distributions of the MOSs and PSNRs are illustrated in Fig.~\ref{mos}, from which it can be observed that the MOSs span from 26.3 to 81.3, covering a wide range of quality levels and substantially spanning the quality spectrum. Therefore, the proposed benchmark epitomizes reasonable perceptual quality separation identifiability.
\subsection{Analyses of Subjective Quality}

\begin{figure}[t]
\centering
\includegraphics[scale=0.34]{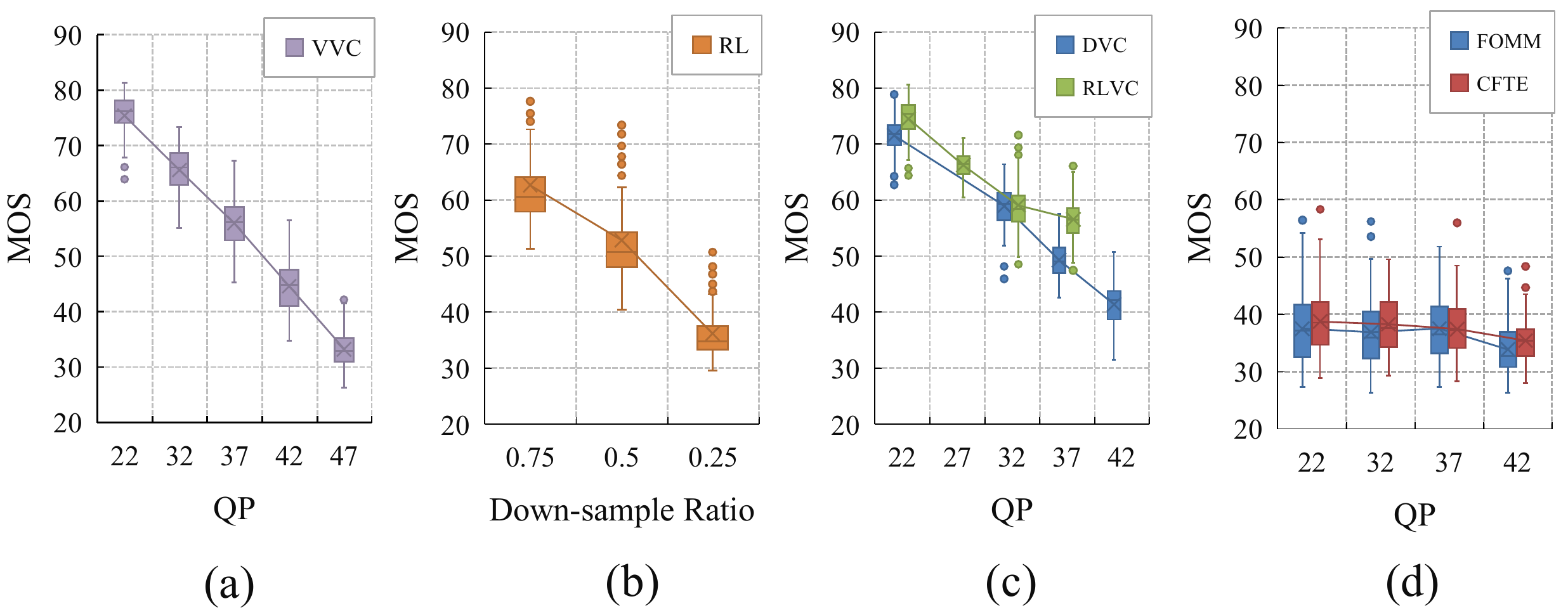}
\caption{Illustration of the MOS variations for different QPs. From bottom to top, the three horizontal lines of each box denote the lower quartile, median, and upper quartile of each subset, respectively. The boundary of the lower whisker is the minimum value, and the upper whisker boundary is the maxim value. The x notion denotes the mean value of each subset. All other observed data points outside the boundary of the whiskers are plotted as outliers. (a) VVC; (b) RL; (c)End-to-end coding methods; (d) Generative coding methods.}
\label{interqpmos}
\end{figure}

\begin{figure}[tbp]
\centering
\includegraphics[scale=0.32]{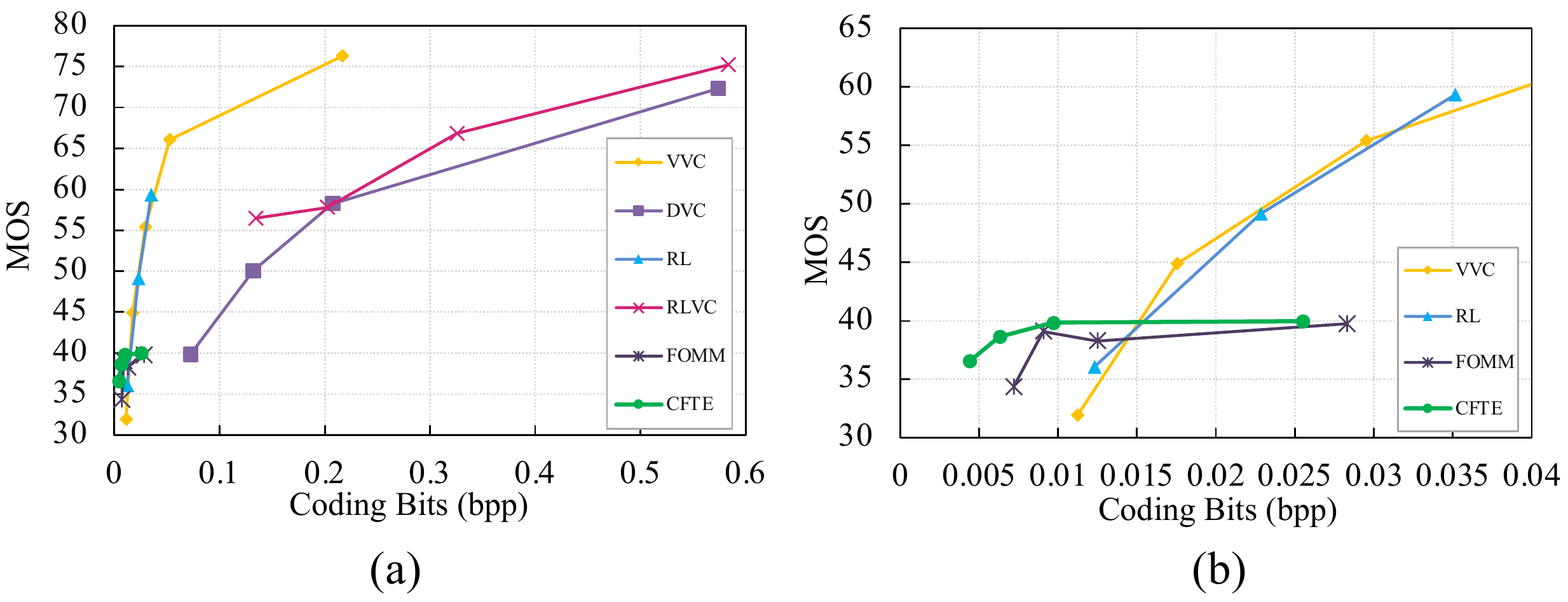}
\caption{The RD performance where the distortion is evaluated in terms of MOS.
(a) The RD curve over the entire bit-rate range. (b) The RD curve in the low bit-rate range.}
\label{rmcurve}
\end{figure}

The MOS distributions for different QPs are shown in Fig.~\ref{interqpmos}, from which, we can observe the performance of different compression models present different sensitivities when the QP increases.
This is not surprising as in FOMM and CFTE, the quality of the first frame controlled by QP is propagated to the subsequent frames by the generative models. As such, the distortions caused by generative models may play more dominant roles. Moreover, recall that the principle in generating compressed videos is ensuring the overlap in subjective quality. As shown in Fig.~\ref{interqpmos}, there is a very apparent quality level overlap among different compression methods, though for generative compression methods the quality levels only span at the low range due to the unique low bitrate coding specialty. 

In Fig.~\ref{rmcurve}, the RD performance regarding the MOS values is illustrated. The bitrate and MOS values of a particular video coding codec are computed by the average values over the compressed videos generated from quarter randomly-sampled compressed video sequences at a specific distortion level. The RD curve can reflect the coding performance faithfully when the quality is measured by the MOS. It can be observed that at the low bitrate range, the generative coding methods are significantly better than the other codecs, which has also been repeatedly observed in \cite{fomm,cfte}. Moreover, with the increase of the bitrate, the visual quality of the generative methods saturates because the distortions from generative models dominate the visual quality. It is also apparent that the performance of RL and VVC is also quite close, as both rely on VVC as the codec. In addition, E2E methods deliver high-quality compressed videos at the expense of a higher bitrate. However, it is worth mentioning that the goal of this work is not comparing different codecs but focusing on evaluating the distortions originating from these codecs instead. 

\section{Performance of Objective Quality Assessment Methods}\label{sec:obj-qa}

We employ eight non-blind image/video quality assessment methods to benchmark the proposed CFVQA database with their quality prediction performance, which are popular measures in the video compression area. In specific, five FR-IQA algorithms, including PSNR~\cite{psnr}, SSIM~\cite{ssim}, Multi-scale Structural Similarity (MS-SSIM)~\cite{msssim}, DISTS~\cite{ding2020image}, and LPIPS~\cite{lpips} are selected, among which DISTS and LPIPS are two data-driven algorithms based on deep neural networks. The quality score for each video is computed on the average quality scores of all frames. Besides, three popular non-blind VQA algorithms STRRED~\cite{strred}, VMAF~\cite{vmaf} and MOVIE~\cite{movie} are also involved. The performance is evaluated by Pearson Linear Correlation Coefficient (PLCC), Spearman Rank Correlation coefficient (SRCC), Kendall Rank Correlation Coefficient (KRCC), and Root-Mean-Square Error (RMSE). Before computing the PLCC and RMSE, objective quality scores are remapped by fitting the non-linear logistic regression function:
\begin{equation}\label{5fitting}
f(x)=\beta_{1}\left(\frac{1}{2}-\frac{1}{1+e^{\beta_{2}\left(x-\beta_{3}\right)}}\right)+\beta_{4} x+\beta_{5},
\end{equation}
where $\beta_{1}$ to $\beta_5$ are the fitting parameters.

In Table~\ref{tab:obj-all}, the quality prediction performance is reported on the entire database and each compression codec subset. It is apparent that STRRED performs the best on four out of six compression subsets, except for RL and RLVC, on which the best performers are VMAF and MOVIE, respectively. The best performer on the entire database is VMAF. 
For the videos compressed by generative models, STRRED and DISTS achieve promising performance in terms of SRCC, but are still far from satisfactory (0.5628 maximum only). The reason is that the generative coding methods produce high-semantic-level distortions, like contour and shape, which HVS is sensitive to. However, current objective measures focus overly on low-level statistical feature mismatches. It typically leads to overestimated quality degradation levels on the GAN-relative distortions, leading to lower correlations with MOSs. On the other side, MS-SSIM also delivers better performance when compared with SSIM either on the entire database scale or each video subset, revealing the superiority of the quality estimation in a multi-scale way. 

\begin{table*}[t]
  \centering
  \caption{Performance of the models on the proposed database and sub-collections. Larger PLCC, SRCC, and KRCC values with smaller RMSE values indicate better performance. In each column, the best performance of each category is boldfaced.}
   \setlength{\tabcolsep}{1.5mm}{\begin{tabular}{cccccccccccccc}
    \toprule
    \multirow{2}{*}{Method} & \multicolumn{6}{c}{SRCC$\uparrow$ for each subset}        & \multicolumn{3}{c}{SRCC$\uparrow$ for each subset} & \multicolumn{4}{c}{Overall} \\
\cmidrule{2-14}     & \multicolumn{1}{c}{VVC} & \multicolumn{1}{c}{RL} & \multicolumn{1}{c}{DVC} & \multicolumn{1}{c}{RLVC} & \multicolumn{1}{c}{FOMM} & \multicolumn{1}{c}{CFTE} & \multicolumn{1}{c}{Traditional} & \multicolumn{1}{c}{End2end} & \multicolumn{1}{c}{Generative} & \multicolumn{1}{c}{PLCC$\uparrow$} & \multicolumn{1}{c}{SRCC$\uparrow$} & \multicolumn{1}{c}{KRCC$\uparrow$} & \multicolumn{1}{c}{RMSE$\downarrow$}  \\
    \midrule
    PSNR  & 0.8997 & \multicolumn{1}{l}{0.8189} & 0.8147 & 0.6375 & 0.4636 & 0.4275 & 0.8794 & 0.7977 & 0.4112 & 0.9017 & 0.8749 & 0.6782 & 6.3026  \\
    SSIM  & 0.8400  & \multicolumn{1}{l}{0.7514} & 0.7416 & 0.5855 & 0.2980 & 0.2845 & 0.8153 & 0.7411 & 0.2669 & 0.8742 & 0.8477 & 0.6420 & 7.0766  \\
    MS-SSIM & 0.9241 & \multicolumn{1}{l}{0.8230} & 0.8700  & 0.7400  & 0.5295 & 0.4808 & 0.8919 & 0.8636 & 0.4408 & 0.9152 & 0.8873 & 0.6975 & 5.8738  \\
    DISTS & 0.9227 & \multicolumn{1}{l}{0.8861} & 0.8805 & 0.7864 & 0.5312 & 0.5542 & 0.9034 & 0.8751 & 0.5449 & 0.8377 & 0.8353 & 0.6420 & 14.5737  \\
    LPIPS & 0.8836 & 0.7930 & 0.8108 & 0.7089 & 0.5218 & 0.5317 & 0.8552 & 0.8196 & 0.5008 & 0.8625 & 0.8644 & 0.6689 & 14.5737 \\
    MOVIE  & 0.9189 & 0.8787 & 0.8798 & \textbf{0.7904} & 0.2363 & 0.1051 & 0.9036 & 0.8753 & 0.1529 &0.7069 & 0.8766 & 0.6854 & 14.5737\\
    STRRED & \textbf{0.9499} & 0.7957 & \textbf{0.9233} & 0.7689 & \textbf{0.5477} & \textbf{0.6010} & 0.8653 & \textbf{0.8960} & \textbf{0.5628} & 0.9117 & 0.8902 & 0.7121 & 5.9862 \\
    VMAF  & 0.9476 & \textbf{0.9271} & 0.9046 & 0.7896 & 0.5472 & 0.4905 & \textbf{0.9443} & 0.8890 & 0.4737 & \textbf{0.9211} & \textbf{0.8984} & \textbf{0.7157} & \textbf{5.3166} \\
    \bottomrule
    \end{tabular}%
    }
  \label{tab:obj-all}%
\end{table*}%

\begin{figure}[t]
\centering
\includegraphics[scale=0.34]{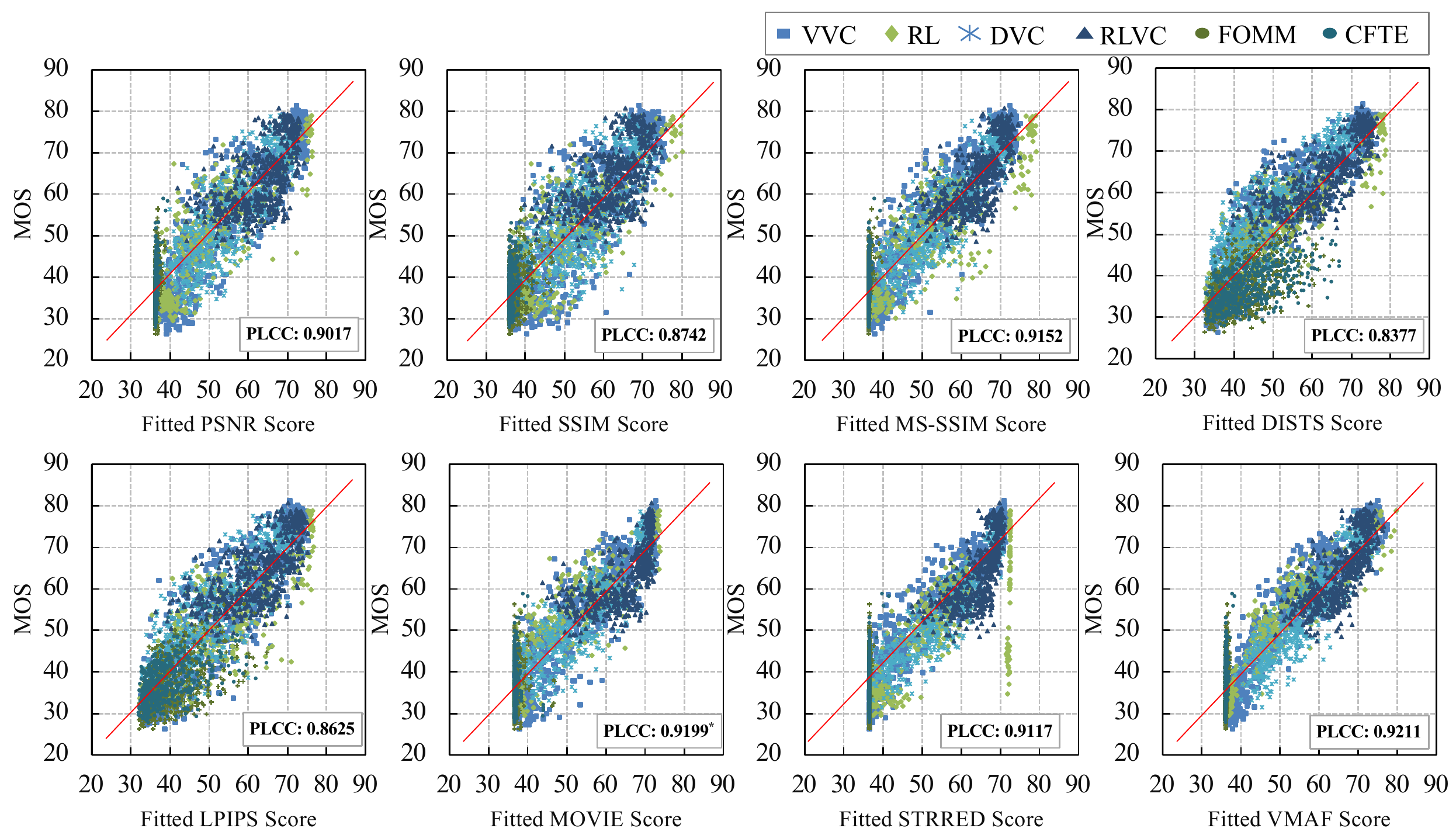}
\caption{Scatter plots of the objective quality scores and MOSs of all compressed face videos in the database. The more diagonally the data points are clustered, the better correlated between objective quality scores and MOSs.}
\label{scatter-all}
\end{figure}
Fig. \ref{scatter-all} shows the scatter plots between the predicted quality of the utilized objective quality measurement and MOSs. These results exhibit that there are still major challenges in the effective quality evaluation of compressed face videos.   It is not surprising to see that the videos from the traditional codec are relatively effortless to handle, while the performance of the E2E encoded videos is lower, potentially due to the flickering artifacts involved, implying that a more effective evaluation for temporal distortions is in need. Overall, the proposed dataset opens up new space for exploring more effective quality assessment models and sheds light on the design of future video compression algorithms. 

\section{When Face Prior Meets Temporal Prior: A Generalized Video Quality Assessment Model for Compressed Face Videos}\label{proposedvqa}
To obtain the compressed video quality objectively, we propose a novel  \textbf{FA}ce \textbf{V}ide\textbf{O} Intege\textbf{R}ity index  (named \textbf{FAVOR}) as an baseline measure for the CFVQA database. The designed philosophy meets both the face prior and temporal prior which all play important roles in human quality perception. Our model FAVOR which achieves promising performance on the proposed dataset and is free from any model retraining or fine-tuning, can serve as the quality measure in the future development of face video compression. 


\subsection{Framework}
\begin{figure}[!tbp]
\centering
\includegraphics[scale=0.48]{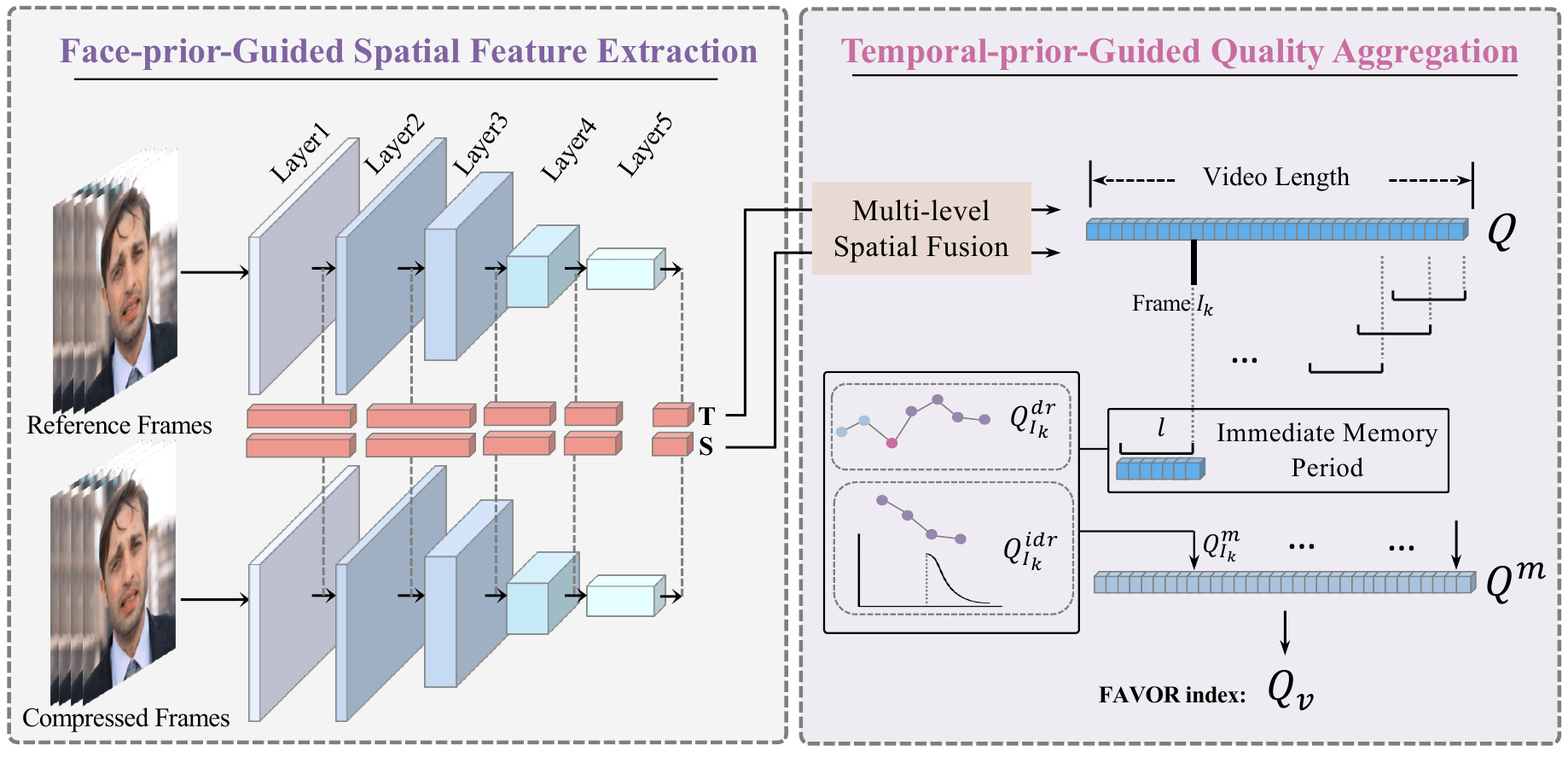}
\caption{The framework of the proposed FR-VQA method FAVOR. It contains the face-prior-guided spatial feature extraction module and the temporal-prior-guided quality fusion module which are sequentially connected. }
\label{framework}
\end{figure}

Following the previous analysis, we basically consider three key elements regarding the perceptual anomalousness in compressed face videos: cognitive attributes of face content, hybrid distortions on low and high semantic levels, and niche temporal artifacts. The flowchart of our FAVOR is presented in Fig.~\ref{framework}, which consists of two stages: (1) The \textit{frame-level quality evaluation} with the face content prior. (2) The \textit{quality aggregation} with the temporal memory prior. In particular, we first adopt a Siamese network to extract quality-aware features for each frame in the reference and compressed face videos. The backbone of Siamese is a face recognition framework ResNet50 which has been fully trained on the MS-Celeb-1M (MS1M) database~\cite{deng2019arcface}.  Herein, the backbone endows the extracted features with high relevance with HVS mechanism for face videos, which can be deemed as a strong spatial prior. Based upon the predicted frame-level quality, we further propose a human memory inspired quality aggregation module and finally achieve a promising prediction result for the whole video quality. The design details are elaborated as follows.

\noindent\textbf{Frame-level Quality Prediction}. 
To capture the quality of each frame, the similarity between the feature maps of the reference frame and distorted frame is evaluated. In particular, we adopt the statistics including the mean $\mu$ and standard deviation (std) $\delta$ of each feature map for the quality evaluation. As indicated in ~\cite{ding2020image}, the mean $\mu$ and std $\delta$ of the deep feature maps are highly correlated with image texture and structure, such that the quality can be effectively estimated by comparing the two terms, 
\begin{equation}
\begin{split}
    t_{i,j} =\frac{2{\mu_{i,j}^r}{\mu_{i,j}^d}+\tau}{(\mu_{i,j}^r)^2+(\mu_{i,j}^d)^2+\tau}, \\
    s_{i,j} =\frac{2{\delta_{i,j}^{rd}}+\tau}{(\delta_{i,j}^r)^2+(\delta_{i,j}^d)^2+\tau},
\end{split}\label{st}
\end{equation}
where $t_{i,j}$ and $s_{i,j}$ denote the texture and structure similarity of the feature map of the \(j\mbox{-}th\) channel in the \(i\mbox{-}th\) stage. The $\mu_{i,j}^r$  and $\delta_{i,j}^r$  are the mean value and std value of the \(j\mbox{-}th\) channel  in the \(i\mbox{-}th\) stage feature map of the reference frame, respectively. The $\mu_{i,j}^d$  and $\delta_{i,j}^d$ have the same meaning with  $\mu_{i,j}^r$  and $\delta_{i,j}^r$, except they are extracted from the corresponding distorted frame. The $\delta_{i,j}^{rd}$ denotes the covariance between the reference and distorted features. The $\tau$ is a small positive offset to avoid zero division. Based upon Eqn.~\eqref{st}, for the $k\mbox{-}th$ frame (denoted as $I_k$) of the test video, its frame-level quality (denoted as $Q_{I_k}$) can be derived by,
\begin{equation}
    Q_{I_k} = \sum_{i=1}^{M}\sum_{j=1}^{N_i}\left(\alpha_{ij}{t_{i,j}}+\beta_{ij}{s_{i,j}}\right),
\end{equation}
where $\alpha_{ij}$ and $\beta_{ij}$ are two weights set defaultly as in~\cite{ding2020image}. $M$ is the total layer number of the features extracted by the backbone and is set to 5, and $N_i$ is the channel number of the $i\mbox{-}th$ feature layer.

\noindent\textbf{Frame-level Quality Aggregation}. In this part, we aim to aggregate all the frame-level quality for the final video quality estimation. The  aggregation philosophy is in line with the human memory effect during quality perception. The hysteresis is that the poor quality in the past frames would cause negative effects on current frame quality~\cite{hysteresis}. With such prior, we model the memory effect within the immediate memory period with the worst quality and refine it into two parts: the \textit{direct effect} and the \textit{indirect effect}. In detail, for the $k\mbox{-}th$ frame $I_k$, its quality is refined by involving the two types of memory effects as follows,
\begin{equation}
Q^m_{I_k} =\left\{
    \begin{aligned}
    Q_{I_k} & , & k \textless l, \\
    \gamma Q^{dr}_{I_k}  +(1-\gamma){Q^{idr}_{I_k} } & , & k \geq l.
    \end{aligned}
    \right.
\label{eqqm}
\end{equation}
The $Q^m_{I_k}$ is the refined quality and the $Q^{dr}_{I_k}$ and $Q^{idr}_{I_k}$ are the direct affected element and indirect affected element, by the past worst quality frame. The $\gamma$ adjusts the levels of the two types of effects. In detail, as shown in Fig. \ref{framework}, for frame $I_k$, we set the period of the immediate memory effect as the past $l$ frames \textit{aka} $\{{I_{(k-l+1)}}, {I_{{(k-l+2)}}}, ..., {I_k} \}$. Suppose  $k \geq l$, and the $p\mbox{-}th$ frame in the interval is with the worst quality, \textit{i.e.}, $Q_{I_{(k-l+p)}} = \mathop{\min}\{Q_{I_{(k-l+1)}}, Q_{I_{{(k-l+2)}}}, ... , Q_{I_k} \}$, the direct effect element $Q^{dr}_{I_k}$ is defined by,
\begin{equation}
\label{hysteresis}
    Q^{dr}_{I_k} = Q_{I_{(k-l+p)}}. 
\end{equation}
This element indicates the worst quality score of the past $l$ frames plays a direct role in the current frame. {In addition to the direct effect, the appearance of the worst frame also persuades the subject to pay much attention  to its successive frames, which is deemed as the indirect effect of the worst-quality frame. We model the  indirect effect as follows, }
 \begin{equation}
    Q^{idr}_{I_k} =\sum_{z=1}^{l-p}{w_z}\cdot{Q_{I_{(k-l+p+z)} }},
\end{equation}
where $w_z$ is the attention weight of the quality score of the $z\mbox{-}th$ frame in the successive ones. In particular, $w_z=f(z^r)$ and  \(f(\cdot)\) is the descending half of Gaussian function where $z^r$ is the quality rank of the $z\mbox{-}th$ frame in the memory period (lower quality rank means lower quality score).  The weighting philosophy lies in that the lower the quality, the negative effect is higher on the  quality evaluation of frame ${I_k}$. Specifically, the $w$s satisfy  ${\sum_{z=1}^{l-p}w_z}=1$. After refining the quality of each frame ${I_k}$ with memory effect, the quality aggregation thus can be performed more reliably as follows,
\begin{equation}
    Q_{v} = \frac{1}{L}\sum_{k=1}^L{{Q^m_{I_k}}},
\end{equation}
where the $Q_{v}$ is the predicted quality of the compressed face video. $L$ is the frame number of the video.  Herein, a higher $Q_{v}$ indicates a better video quality.

\subsection{Performance Evaluations}
\noindent\textbf{Implementation Details}. Our framework is implemented by PyTorch on a machine equipped woth NVIDIA Geforce RTX 2080Ti GPUs. Following the operation in \cite{ssim} and \cite{ding2020image}, the face videos are resize to 224$\times$ 224 while keeping the original aspect ratio, and input into the DNN backbone in the RGB formation. We adopt five stages of covolutional responses including $7\mbox{-}th$, $16\mbox{-}th$, $45\mbox{-}th$, $52\mbox{-}nd$ convolutional layers and the final output of the pretrained ResNet50 are adopt for embedding comparison, which contains 64, 128, 256, 512, 512 feature channels respectively. The $\gamma$ and $l$ in the Eqn.~\ref{eqqm} are tuned to be set as 0.1 and 4 respectively. 


\noindent\textbf{Quality Prediction Performance}.
We compare the performance of the FAVOR index with other five IQA methods\cite{psnr,ssim,msssim,ding2020image,lpips} and three VQA methods\cite{movie,strred,vmaf}. 
The IQA methods generate frame-level scores and are aggregated averagely to compose the video quality score. The results are presented in Table~\ref{tab:perf-ours}, which indicate that the proposed FAVOR index outperforms the other methods in terms of the PLCC, SRCC, and KRCC. Also, we present the fine-grained quality prediction performance~\cite{zhang2021fine} on each QP point of different compression methods. The results show that the proposed method not only achieves the state-of-the-art performance on 14 out of 24 video collections, but also on the entire CFVQA database, indicating a promising overall and fine-grained quality prediction performance. In particular, for the generative compression collection FOMM and CFTE, FAVOR shows superiority over other methods.
\begin{table*}[!tbp]
  \centering
  \fontsize{7}{8}\selectfont
  \caption{Fine-grained Quality Prediction Performance on Each QP point of Different Compression Codecs and Overall Performance on the Entire CFVQA Database In Terms Of SRCC. The FAVOR Performance is Highlighted in Purple, and the \textbf{Bold} Number Represents the Best Performance.}
    \setlength{\tabcolsep}{1.4mm}{\begin{tabular}{c|cccc|cccc|cccccccc}
    \toprule
    \multicolumn{1}{c}{} & \multicolumn{8}{c|}{Fine-grained SRCC on the Videos Compressed by Traditional Codecs} & \multicolumn{8}{c}{Fine-grained SRCC on the Videos Compressed by E2E Codecs} \\
    \midrule
    \multirow{2}{*}{Metric} & \multicolumn{5}{c|}{VVC}              & \multicolumn{3}{c|}{RL} & \multicolumn{4}{c|}{DVC}      & \multicolumn{4}{c}{RLVC} \\
          & 22    & 32    & 37    & \multicolumn{1}{c}{42} & \multicolumn{1}{c|}{47} & 0.25  & 0.5   & 0.75  & 22    & 32    & 37    & \multicolumn{1}{c|}{42} & 22    & 27    & 32    & 37 \\
    \midrule
    \midrule
    PSNR~\cite{psnr}  & 0.9270 & 0.8799 & 0.8956 & \multicolumn{1}{c}{0.9059} & \multicolumn{1}{c|}{0.9025} & 0.3878 & 0.4807 & 0.3367 & 0.8396 & 0.7852 & 0.7970 & \multicolumn{1}{c|}{0.8319} & 0.7231 & 0.5777 & 0.5718 & 0.7051 \\
    SSIM~\cite{ssim}  & 0.8980 & 0.7974 & 0.8254 & \multicolumn{1}{c}{0.8244} & \multicolumn{1}{c|}{0.8712} & 0.3497 & 0.4641 & 0.2411 & 0.7956 & 0.5718 & 0.7100 & \multicolumn{1}{c|}{0.7612} & 0.6998 & 0.5178 & 0.5142 & 0.6296 \\
    MS-SSIM~\cite{msssim} & 0.9363 & 0.9070 & 0.9385 & \multicolumn{1}{c}{0.9096} & \multicolumn{1}{c|}{\textbf{0.9444}} & 0.4489 & 0.5715 & 0.4616 & 0.8710 & 0.8827 & 0.8659 & \multicolumn{1}{c|}{0.8547} & 0.7567 & 0.7306 & 0.7257 & 0.7857 \\
    DISTS~\cite{ding2020image} & 0.9330 & 0.8883 & \textbf{0.9475} & \multicolumn{1}{c}{0.9321} & \multicolumn{1}{c|}{0.9099} & 0.5060 & 0.5994 & 0.5221 & 0.8754 & 0.8913 & 0.8824 & \multicolumn{1}{c|}{0.8678} & 0.7778 & 0.7731 & \textbf{0.8182} & 0.7859 \\
    LPIPS~\cite{lpips} & 0.8882 & 0.8283 & 0.9130 & \multicolumn{1}{c}{0.9049} & \multicolumn{1}{c|}{0.8981} & 0.4259 & 0.5199 & 0.4542 & 0.7753 & 0.8221 & 0.8312 & \multicolumn{1}{c|}{0.8192} & 0.6882 & 0.6846 & 0.7422 & 0.7284 \\
    MOVIE~\cite{movie} & 0.9349 & 0.9256 & 0.9176 & \multicolumn{1}{c}{0.9289} & \multicolumn{1}{c|}{0.8891} & 0.4122 & 0.4606 & 0.2073 & 0.8662 & 0.8699 & 0.8887 & \multicolumn{1}{c|}{0.8582} & \textbf{0.8423} & 0.7532 & 0.8006 & 0.7836 \\
    STRRED~\cite{strred} & 0.9324 & 0.9238 & 0.9205 & \multicolumn{1}{c}{0.9327} & \multicolumn{1}{c|}{0.9313} & \textbf{0.6084} & 0.6004 & 0.4968 & \textbf{0.9200} & 0.9179 & 0.9247 & \multicolumn{1}{c|}{0.9103} & 0.7726 & \textbf{0.7941} & 0.7361 & 0.7815 \\
    VMAF~\cite{vmaf}  & 0.9352 & 0.9281 & 0.9292 & \multicolumn{1}{c}{0.9271} & \multicolumn{1}{c|}{0.9323} & 0.5011 & 0.5517 & 0.5152 & 0.8997 & 0.9176 & 0.9041 & \multicolumn{1}{c|}{0.9040} & 0.7977 & 0.7763 & 0.7881 & \textbf{0.8133} \\
    \rowcolor{mygreen} \textbf{FAVOR} & \textbf{0.9427} & \textbf{0.9296} & 0.9323 & \multicolumn{1}{c}{\textbf{0.9377}} & \multicolumn{1}{c|}{0.9378} & 0.5656 & \textbf{0.6381} & \textbf{0.6552} & 0.8757 & \textbf{0.9249} & \textbf{0.9232} & \multicolumn{1}{c|}{\textbf{0.9163}} & 0.7296 & 0.7327 & 0.7263 & 0.7351 \\
    \midrule
    \multicolumn{1}{c}{} & \multicolumn{8}{c|}{Fine-grained SRCC on the Videos Compressed by Generative Codecs} & \multicolumn{8}{c}{Performance on the CFVQA Database} \\
    \midrule
    \multirow{2}{*}{Metric} & \multicolumn{4}{c|}{FOMM}     & \multicolumn{4}{c|}{CFTE}     & \multicolumn{8}{c}{Overall} \\
          & 22    & 32    & 37    & 42    & 22    & 32    & 37    & 42    & \multicolumn{2}{c}{PLCC$\uparrow$} & \multicolumn{2}{c}{SRCC$\uparrow$} & \multicolumn{2}{c}{KRCC$\uparrow$} & \multicolumn{2}{c}{RMSE$\downarrow$} \\
    \midrule
    \midrule
    PSNR~\cite{psnr}  & 0.4925 & 0.5105 & 0.4916 & 0.4160 & 0.4403 & 0.4621 & 0.4964 & 0.3490 & \multicolumn{2}{c}{0.9017} & \multicolumn{2}{c}{0.8749} & \multicolumn{2}{c}{0.6782} & \multicolumn{2}{c}{6.3026} \\
    SSIM~\cite{ssim}  & 0.3479 & 0.3732 & 0.3390 & 0.2970 & 0.3149 & 0.3492 & 0.3251 & 0.2154 & \multicolumn{2}{c}{0.8742} & \multicolumn{2}{c}{0.8477} & \multicolumn{2}{c}{0.6420} & \multicolumn{2}{c}{7.0766} \\
    MS-SSIM~\cite{msssim} & 0.5382 & 0.5665 & 0.5477 & 0.5205 & 0.4975 & 0.5139 & 0.5291 & 0.4511 & \multicolumn{2}{c}{0.9152} & \multicolumn{2}{c}{0.8873} & \multicolumn{2}{c}{0.6975} & \multicolumn{2}{c}{5.8738} \\
    DISTS~\cite{ding2020image} & 0.5121 & 0.5663 & 0.5328 & 0.4030 & \textbf{0.6466} & 0.5623 & 0.4904 & 0.4216 & \multicolumn{2}{c}{0.8377} & \multicolumn{2}{c}{0.8353} & \multicolumn{2}{c}{0.6420} & \multicolumn{2}{c}{14.5737} \\
    LPIPS~\cite{lpips} & 0.5456 & 0.5146 & 0.4933 & 0.4134 & 0.5208 & 0.5086 & 0.4898 & 0.4530 & \multicolumn{2}{c}{0.8625} & \multicolumn{2}{c}{0.8644} & \multicolumn{2}{c}{0.6689} & \multicolumn{2}{c}{14.5737} \\
    MOVIE~\cite{movie} & 0.2970 & 0.2755 & 0.2195 & 0.1701 & 0.1277 & 0.1440 & 0.1499 & -0.0017 & \multicolumn{2}{c}{0.7069} & \multicolumn{2}{c}{0.8766} & \multicolumn{2}{c}{0.6854} & \multicolumn{2}{c}{14.5737} \\
    STRRED~\cite{strred} & 0.5699 & 0.5321 & 0.5004 & 0.5014 & 0.6284 & 0.6097 & 0.6125 & 0.5281 & \multicolumn{2}{c}{0.9117} & \multicolumn{2}{c}{0.8902} & \multicolumn{2}{c}{0.7121} & \multicolumn{2}{c}{5.9862} \\
    VMAF~\cite{vmaf}  & 0.5677 & \textbf{0.5752} & 0.5228 & 0.5568 & 0.4967 & 0.4822 & 0.5199 & 0.4467 & \multicolumn{2}{c}{0.9211} & \multicolumn{2}{c}{0.8984} & \multicolumn{2}{c}{0.7157} & \multicolumn{2}{c}{\textbf{5.3160}} \\
    \rowcolor{mygreen}{\textbf{FAVOR}} & \textbf{0.5818} & 0.5289 & \textbf{0.5657} & \textbf{0.5628} & 0.6158 & \textbf{0.6173} & \textbf{0.6154} & \textbf{0.5379} & \multicolumn{2}{c}{\textbf{0.9229}} & \multicolumn{2}{c}{\textbf{0.9060}} & \multicolumn{2}{c}{\textbf{0.7248}} & \multicolumn{2}{c}{5.6117} \\
    \bottomrule
    \end{tabular}%
    }
  \label{tab:perf-ours}%
\end{table*}%


\begin{table}[!tbp]
  \centering
  \fontsize{7}{7}\selectfont
  \caption{Performance of spatial and temporal measures when evaluated in isolation on the CFVQA database. ``T'' denotes the IQA methods aggregated by the proposed temporal aggregation module. }
    {\begin{tabular}{ccccc}
    \toprule
    Method & PLCC$\uparrow$ & SRCC$\uparrow$ &KRCC$\uparrow$ &RMSE$\downarrow$ \\
    \midrule
    average-$Q_{I_k}$ & 0.9068 & 0.8876 & 0.7123 & 5.7441 \\
    hysteresis~\cite{hysteresis}-$Q_{I_k}$ &0.9060 &0.8997   &0.7152   &5.8462   \\
    recency~\cite{recency}-$Q_{I_k}$ &0.9062   &0.8857 &0.6941 &6.1619 \\
    VQPooling~\cite{vqpooling}-$Q_{I_k}$ &0.9001 &0.8792 &0.6887 &6.5915 \\
    percentile~\cite{rimac2009influence}-$Q_{I_k}$    &0.9135 &0.8915 &0.7088 &5.9281\\
    primacy~\cite{recency}-$Q_{I_k}$   &0.8307 &0.7988 &0.6111 &8.1145\\
    variation~\cite{ninassi2009considering}-$Q_{I_k}$ &0.6330 &0.6027 &0.4327 &11.2822\\
    \midrule
    T-PSNR  &0.9038 &0.8803 &0.6912 &6.3011\\
    T-SSIM  &0.8900 &0.8576 &0.6513 &6.8478\\
    T-MS-SSIM    &0.9158  &0.8883    &0.6980 &5.8722\\
    T-DISTS &0.8724 &0.8572 &0.6436 &7.3779\\
    T-LPIPS &0.8755 &0.8689 &0.6803 &8.9137\\
    \midrule
    \rowcolor{mygreen} \textbf{FAVOR} & \textbf{0.9229} & \textbf{0.9060} & \textbf{0.7248} & \textbf{5.6117} \\   
    
    \bottomrule
    \end{tabular}%
    }
  \label{ablation}%
\end{table}%

\noindent\textbf{Ablation Study}. The main components of our method are the face-prior-based quality predictor and the memory-prior-based quality aggregator. To reveal the effectiveness of each component,  extensive ablation studies are conducted. 
\begin{itemize}
    \item  \textit{Study of the frame quality predictor.} In order to investigate the effectiveness of the quality predictor, we ablate it from our model and adopt  other commonly-used IQA-based models for the frame-level quality prediction. The models compared against include PSNR, SSIM, MS-SSIM, DISTS, and LPIPS. The results of this study are presented in Table~\ref{ablation}, which reveals that that all the alternatives lead to a performance drop. This observation indicates that more accurate quality can be obtained by incorporating facial prior information.

 \item  \textit{Study of the quality aggregator.} To study our memory-inspired aggregation strategy, we replace it with various aggregation modules, including the average pooling, temporal hysteresis\cite{hysteresis}, recency effect\cite{recency}, VQPooling~\cite{vqpooling}, percentile\cite{rimac2009influence}, primacy\cite{recency}, and temporal variation\cite{ninassi2009considering}. The results are presented in Table~\ref{ablation}. It can be observed that our aggregation strategy outperforms the alternative strategies by a significant margin, demonstrating the high complementary between our temporal quality aggregator and our spatial quality predictor. Moreover, from the Table, we can observe all the IQA-based models benefit from our quality aggregator, which further proves the high-generalization capability of our quality aggregator. 
\end{itemize}
\begin{figure}[tbp]
\centering
\includegraphics[scale=0.52]{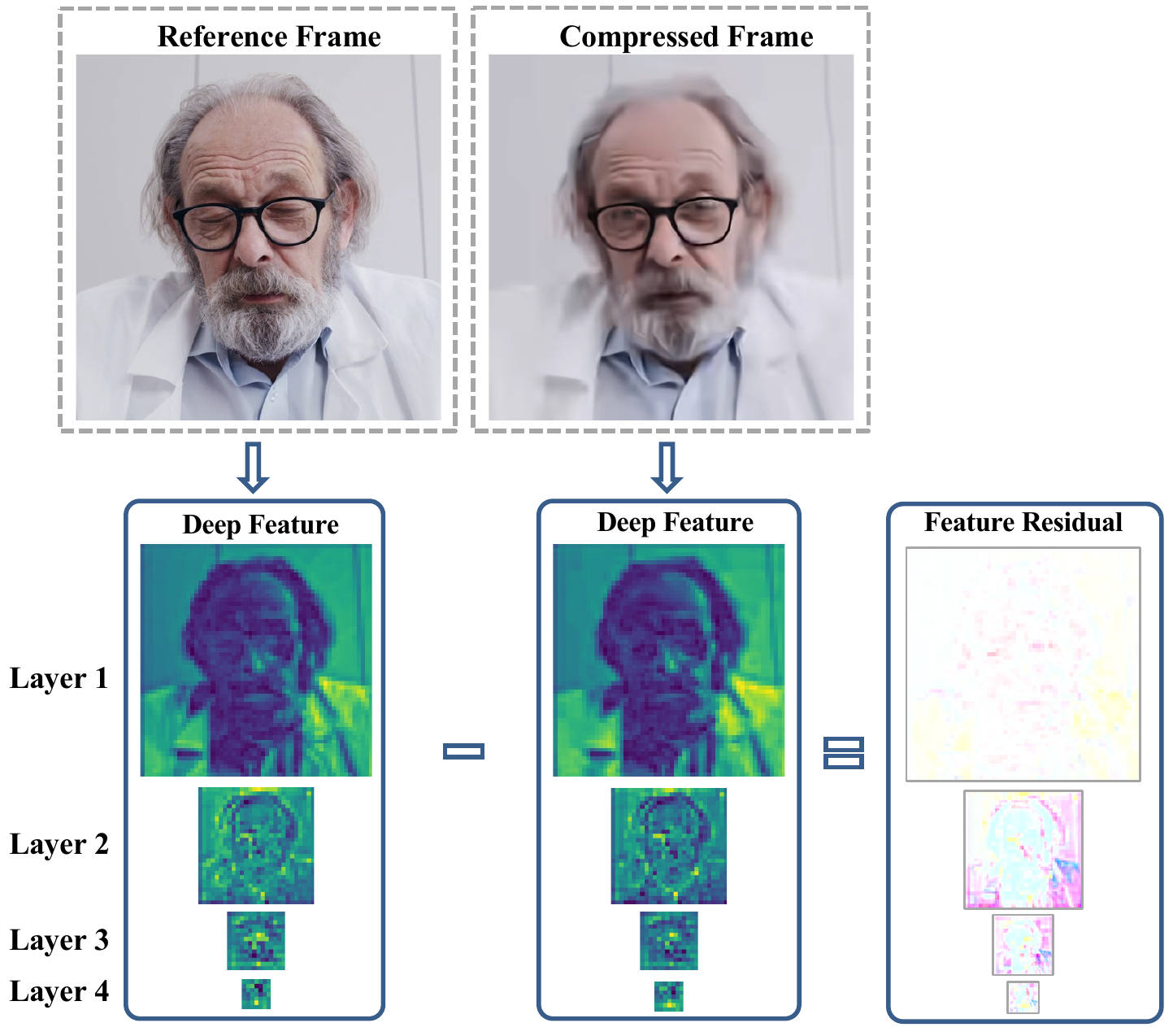}
\caption{Sampled feature maps from the four output layers of the ResNet decomposition of the reference frame and the VVC-compressed frame. Note the output of the fifth layer cannot be visualized due to it is a one-dimensional vector. More examples can be found in the SM.}
\label{featuremaps}
\end{figure}
\noindent\textbf{Feature Visualisation}. In Fig. \ref{featuremaps}, we present the feature maps of both the reference and corresponding compressed frame. The residual maps are further measured to visualize the quality corruption. From the residual maps, we can observe the quality degradation clues exist in different layers and the deeper the layer, the more detectable is the corruption. The reason may lie that the high-level  semantics are severely corrupted after compression.

\section{CONCLUSIONS}\label{sec:conclusion}
In this paper, we have addressed a unique quality assessment problem that is critical for face video compression, coinciding with the accelerated development of artificial intelligence. We systematically study the problem by creating a dataset CFVQA and developing a dedicated FR-VQA algorithm named FAVOR. 
The dataset is featured with large scale (i.e., 103,680 labels with 3,240 distorted videos), multiple distortion types (i.e., traditional, E2E, and generative distortions), and a wide range of distortion levels. The objective VQA measure FAVOR, which particularly relies on spatial and temporal priors, has been validated to be effective without requiring any retraining process with the proposed dataset. We demonstrate the necessity of feature extraction and fusion incorporating the face prior for compressed face video quality evaluation. This study, the first attempt at compressed face video quality assessment, has great potential to facilitate other compression or quality assessment research. For example, the visual quality assessment of semantic compression methods can be thoroughly investigated along this vein. Moreover, more advanced VQA methods that deliver robust performance over different distortion types can be developed for compressed videos as triggered and enabled by this study.


%




\ifCLASSOPTIONcaptionsoff
  \newpage
\fi



%
\bibliographystyle{IEEEtran}
\bibliography{refs}



%




\end{document}